\documentclass[aps,prb,floats, 
graphicx,euscript,amsmath, amssymb, twocolumn, 10pt]{revtex4}
\usepackage[dvips]{graphicx}
\usepackage[dvips]{color}
\begin{document}
\title{Two channel orbital Kondo effect in quantum dot with SO(n) symmetry}
\author{T. Kuzmenko$^1$, K. Kikoin$^2$ and Y. Avishai$^{1,3}$}
\affiliation{
$^1$Department of physics, Ben Gurion University of the Negev, Beer
Sheva 84105, Israel \\
$^2$Raymond and Beverly Sackler Faculty of Exact
Sciences, School of Physics and Astronomy, Tel Aviv University, 69978 Tel Aviv, Israel\\
$^3$ Department of Physics, Hong Kong University of Science and Technology,
Clear Water Bay, Kowloon, Hong Kong}
\begin{abstract}
 A scenario for the formation of non-Fermi-liquid
(NFL) Kondo effect (KE) with spin variable enumerating Kondo channels
is suggested and worked out. In doubly occupied symmetric triple quantum dot
within parallel geometry, the NFL low-energy
regime arises provided the device possesses both source-drain and
left-right parity. Kondo screening follows a multistage
renormalization group mechanism: reduction of the energy scale is
accompanied by the change of the relevant symmetry group from
$SO(8)$ to $SO(5)$. At low energy, three phases compete: 1)  under-screening spin
triplet  (conventional) KE; 2) spin singlet
potential scattering;  3)  NFL phase where
the roles of spin and orbital degrees of freedom are swapped.

\end{abstract}
\maketitle

\section{Introduction}

The physics of two-channel Kondo effect (2CKE) that is marked by
non-fermi liquid (NFL) behavior at low temperatures has been with us
for more than three decades.\cite{NoBla80}. Complex quantum dots
(CQD) coupled to metallic leads via several channels may be
considered as candidates for possible realization of multichannel
KE. The simplest object of this kind is a double quantum dot coupled
to two source and two lead
electrodes.\cite{Hoza02,Eto05,Lim06,Trocha10} The generic $SU(4)$
symmetry of this device is realized by two spin 1/2 projections and
two orbital states of an electron in a double quantum dot. This
configuration paves the way to observation of the $SU(4)$ Kondo
effect for four-fold degenerate ground state or at least for the
orbital $SU(2)$ KE realized only by orbital degrees. It should be
noted that unavoidable interference between different channels
emerging in cotunneling process through CQD turns the $SU(4)$  fixed
point to be unstable.\cite{Lim06} As a result of this inteference
one of the two channels becomes dominant at low temperatures,
resulting in an $SU(4)\rightarrow SU(2)$ crossover
 (see Ref. \onlinecite{Tetta12} and references therein for
experimental realization of these effects in carbon nanotubes and
multivalley silicon quantum dots).

 In order to realize 2CKE for odd occupation in double quantum
dot structures, one more tunneling channel should be involved in
electron tunneling processes. However the NFL regime is still
elusive due to the same channel anisotropy emerging from
inter-channel cotunneling processes. To remedy this instability, a
suppression of interchannel cotunneling is attempted, using special
design of the CQD. Apparently, the most successful attempt is
realized in double quantum dot (DQD),\cite{Potok07} where the
interference is suppressed by Coulomb blockade. Another design which
allows one to (at least) approach the elusive two-channel fixed
point was suggested in Ref.~\onlinecite{KKA03}, based on a structure
composed of
  triple quantum dot (TQD) in a serial
 geometry (Fig.~\ref{fig_1}, left panel).  Here, the strong Coulomb blockade in the central dot
minimizes (but does not completely eliminate) interchannel
interference.

Yet another approach for achieving NFL two-channel Kondo regime
is to swap the roles of charge and spin variables, i.e., to treat
orbital or charge fluctuations as pseudospin variables causing
 Kondo screening, whereas spin projection quantum numbers
 serve as different channels.\cite{Kondo76,WlaZa83,MuFi96,MuFi97,CoZa98}
The proposal was based on the idea that the orbital degrees of
freedom  of heavy spinless particles in a two-well or three-well
potential trap may be converted into pseudospin variables, and the
latter may play part of the source of Kondo screening, while the
spin projections of conduction electrons enumerate the screening
channels. This idea was subject to
 criticism.\cite{KaPro86,Aleiner01,AlContr02} In these systems,
pseudospin-flip is a generic tunneling process with a characteristic
(long) time $t_{\rm tun}$, unlike real spin flip processes which are
practically instantaneous. As a result the ultraviolet cutoff for
Kondo effect is the energy $\sim \hbar/t_{\rm tun}$. This energy is
of the order of the distance to the next excited level in two-well
potential. The latter interval is much smaller than the energy scale
$\varepsilon_F$ for "light" electrons. As a result the energy
interval available for the formation of the logarithmic singularity
is too narrow and the resulting Kondo temperature is very small
$T_K\ll \Delta$  where $\Delta$ is the depth of the occupied level
in the well relative $\varepsilon_F.$ Thus the strong coupling
regime remains in fact unattainable as far as two-level systems in
heavy particles serve as pseudospin. Some theoretical
counter-arguments have been offered later,\cite{Zar05} whereas
features of NFL behavior were found in the electrical resistivity of
glassy ThAsSe single crystal.\cite{Cich} Thus, the question of
whether 2CKE can be unambiguously realized in two-level systems is
still under discussion (see also Ref. \onlinecite{Delft}).

Natural and artificial nano-objects provide their own mechanisms of
two-channel Kondo tunneling assisted by pseudospin excitations. One
such mechanism was proposed for a "quantum box" connected to a lead
by a single-mode point contact. \cite{Matv95,LebShil01} In this case
the relevant operator which logarithmically scales the crossover
from the high temperature to the low-temperature region is the
capacitance $C(T)\sim \ln(T/T_K)$. In another model with $SU(3)$
dynamical symmetry the excited state with parity-degenerate
rotational levels $(m = \pm 1)$ may cross the $m=0$ level due to
interaction with a bath and thus become the source of orbital KE
with spin playing the part of a tunneling channel. \cite{Arnold07}
One more possibility of realizing 2CKE was discussed
recently\cite{Law10} for a quantum dot coupled to two helical edge
states of a 2D topological insulator.

\section{Model}

\begin{figure}[h]\begin{center}
  \includegraphics[width=8.5cm,angle=0]{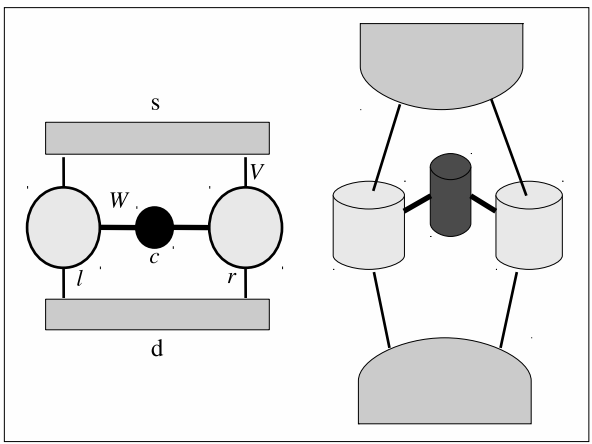}
  \caption{Axially symmetric triple quantum dot in planar (left panel) and vertical (right panel)
geometries.}\label{fig_1}
\end{center}
\end{figure}

In this work we offer a relatively simple realization of
over-screened orbital Kondo effect where the two-channel regime is
realized by two spin projections. The proposed device  is composed
of a doubly occupied TQD in contact with two terminals within
parallel geometry \cite{KKA03}. It will be shown that the present
model is free of the shortcomings
 pointed out in Refs. \onlinecite{KaPro86,Aleiner01,AlContr02}
because the role of higher excited levels is completely different.
Experimentally, this configuration may be realized in triangular
arrangement of vertical dots \cite{Amaha09} (see Fig~\ref{fig_1},
right panel).
The starting point is the usual Anderson-like tunneling
Hamiltonian,
\begin{equation}
H = H_d + H_b + H_{db}
\end{equation}
 where
 \begin{eqnarray}
 &&H_d =\sum_{\lambda=\sigma,\Lambda} E_\lambda |\lambda\rangle\langle\lambda| \nonumber \\
 &&H_b = \sum_{i=l,r}\sum_{k\sigma}\sum_{a=e,o} \varepsilon_k
 c^\dag_{iak\sigma}c^{}_{iak\sigma} \\
 &&H_{\rm tun} = \sqrt{2}\sum_{ik\sigma\Lambda}\left(\bar\sigma V_\Lambda c^{\dag}_{iek\sigma}|\bar\sigma\rangle
\langle\Lambda| + {\rm H.c.}\right).  \nonumber
\end{eqnarray}
We consider the configuration possessing both left-right ($l$-$r$) and
source-drain ($s$-$d$)
symmetry. The operators $c^{\dag}_{iak\sigma}$=$
(c^{\dag}_{isk\sigma}\pm c^{\dag}_{idk\sigma})/\sqrt{2}$ are even
and odd combinations of source and drain electron operators. The latter combinations
do not enter $H_{\rm tun}$.\cite{Eto05,Dynsym}  The
TQD is doubly occupied in the ground state, and only singly
occupied states are involved in cotunneling processes. The
corresponding eigenstates of the isolated TQD $|\lambda\rangle$
are denoted as $|\sigma\rangle$  and $|\Lambda\rangle$ for the TQD
occupied by $N$=1 (spin doublets), and $N$=2 (spin triplets and
singlets) electrons respectively.

The low-energy spectrum $E_\Lambda$ of the isolated doubly occupied
TQD consists of two singlets $E_{Si}$ and two spin triplets
$E_{Ti}$ ($i= g,u$ for even and odd combinations of $l$ (left) and $r$(right)
states)
\begin{eqnarray}
&&E_{S_g}=\tilde{\varepsilon}-\frac{2W^2}{\Delta
}-\frac{2W^2}{\Delta+Q}-\frac{4W^2}{\varepsilon_c+Q_{c}-Q_{ic}-\varepsilon},\nonumber\\
&&E_{T_u}=\tilde{\varepsilon}-\frac{2W^2}{\Delta
},\nonumber\\
&&E_{S_u}=\tilde{\varepsilon}-\frac{2W^2}{\Delta
+Q},\nonumber\\
&&E_{T_g}=\tilde{\varepsilon},\label{En}
\end{eqnarray}
with $\Delta=\varepsilon-\varepsilon_c-Q_{ic}$ (see Appendix A for
detailed calculation of these eigenstates and the corresponding
eigenfunctions $|\Lambda\rangle$ of the Hamiltonian $H_d$).

\begin{figure}[h]\begin{center}
  \includegraphics[width=8.5cm,angle=0]{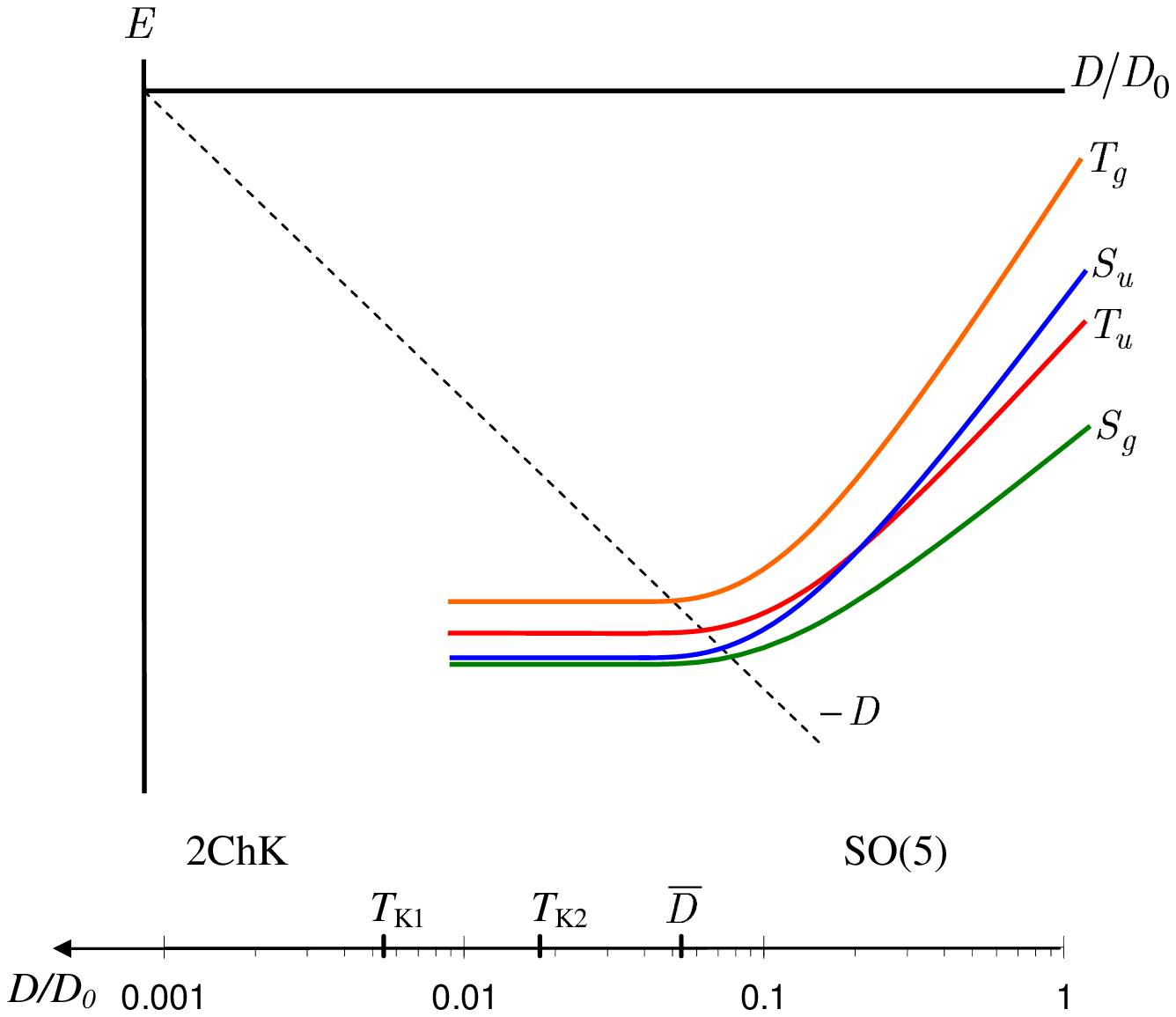}
  \caption{(Color online) Upper panel: Energy levels in doubly occupied TQD renormalized within a
scaling procedure. Lower panel: Logarithmic energy scale for a
crossover from SO(5) to two-channel KE as a function of scaling
parameter $\eta=\ln D_0/D$.}
\vspace{-0.2in}
  \label{fig_2}
\end{center}
\end{figure}

Within the energy scale of the bandwidth $D$ (exceeding the width
of this multiplet) the spectrum of the isolated TQD
 is characterized by $SO(8)$
dynamical symmetry.  Dynamical symmetry group characterizes the symmetry of the interlevel
transitions rather than the symmetry of the Hamiltonian. These transitions involve not only the states
belonging to the same irreducible representation characterizing the symmetry of the Hamiltonian but
also the processes connecting non-degenerate states belonging to different group representations  (see \cite{Dynsym} for a
regular description of dynamical symmetries).   Mathematically, dynamical symmetry group of any quantum mechanical
 system is a group such that all the states of interest are contained in a single irreducible
representation of the group. In our case these states are two singlets and two triplets involved in
the scaling renormalization (Fig. 2). The dynamical symmetry group is formed by
linear combinations of the operators
$X^{\lambda\lambda'}=|\lambda\rangle\langle\lambda'|$ generating
the spectrum of the Hamiltonian $H_d$.   Haldane
renormalization \cite{Dynsym,KA01,KKA04,Hald78}
 implies that
all the levels flow downwards
with scaling invariants,
\begin{equation}\label{halda}
E^*_\Lambda = E_\Lambda(D)
-\pi^{-1}\Gamma_\Lambda\ln(D/\Gamma_\Lambda).
\end{equation}
Here $\Gamma_\Lambda = \pi \rho_0 |V_{\Lambda}|^2$ is the tunneling
rate for the state $|\Lambda\rangle$, $\ln (D/D_0) \equiv \eta$ is
the scaling variable and $D_0$ is the conduction bandwidth $-D_0/2
\le \varepsilon_k \le D_0/2$. The renormalization rates
$\Gamma_\Lambda/\pi$ depend crucially on $\Lambda$, and the scaling
trajectories $E_\Lambda(D)$ intersect as in Fig. \ref{fig_2} due to
the inequality $\Gamma_{T_g} > \Gamma_{S_u}>
\Gamma_{T_u}>\Gamma_{S_g}$ (see Appendix \ref{epsaps} for details).
Various scenarios of multistage Kondo effect are possible, including
that illustrated by the flow diagram of Fig. \ref{fig_2}, where the
level $E_{S_u}$ crosses level $E_{T_u}$ \textit{before} the level
$E_{T_g}$ "overtakes" both of them.\cite{foot1} There is a window of
input parameters, where the orbital KE emerges due to nearly
degenerate orbital doublet/spin singlet forming as a result of
Haldane flow at $D\approx \bar D$, where charge fluctuations are
frozen. At this stage, an application of the Schrieffer-Wolff (SW)
transformation generates the effective spin Hamiltonian for $N=2$
channels, and further RG transformation follows Anderson's poor-man
scaling procedure for renormalization of the exchange
constants.\cite{And70}

\section{From SO(5) spin Kondo effect to two-channel orbital Kondo effect}

The full $SO(8)$ spin Hamiltonian $H_{\rm SW}$ is derived in Ref.
\onlinecite{KKA04}.  From Fig. \ref{fig_2} we conclude that the
poor-man scaling procedure should take into account the evolution of
dynamical symmetry along the chain $SO(8)\to SO(5)\to {\rm orbital}~
SU(2)$. To illustrate the key points of transformation from spin KE
to orbital KE, we consider the simplified picture, where only one of
the two triplets is taken into account, i.e., write down the spin
Hamiltonian pertaining to the part relevant to the KE for the
$SO(5)$ multiplet composed of one triplet $T_u$ and two singlets
$S_u$ and $S_g$.  When the two singlets $|S_{u}\rangle, |S_g
\rangle$ and the triplet $|T_{u},\mu\rangle$ (where
 $\mu= 1,0,\bar 1$ are the projections of the spin $S=1$ along a given axis), are almost
degenerate, the effective low-energy Hamiltonian can be written as
a sum of spin and orbital parts:
\begin{eqnarray}
H_{\rm eff}&=&H_{\rm spin}+H_{\rm orb}.\label{sp-orb}
\end{eqnarray}
The first term $H_{\rm spin}$ is generated at the first stage of
full Haldane-Anderson scaling procedure. The second term $H_{\rm
orb}$ arises only provided the two lowest states in the
renormalized Hamiltonian $H_d$ are spin singlets (see below).
$H_{\rm spin}$ contains twelve coupling constants
(\ref{sym52-tlr}) (see Appendix B for this derivation), however
only three of them are relevant, and the Kondo temperature can be
 identified from the reduced effective Hamiltonian
\begin{eqnarray}
&&H_{\rm spin} =\sum_{\mu}{\bar
E}_{T_u}X^{T_u\mu,T_u\mu}+\sum_{\eta=u,g}{\bar
E}_{S_\eta}X^{S_\eta S_\eta} +
\label{sym-so5}\\
&& J_{1} {\bf S}_{u}\cdot {\bf s}_{uu} +J_{2}{\bf R}_{u}\cdot {\bf
s}_{uu}+J_{3}({\bf R}_{ug}^{(1)}\cdot {\bf s}_{gu}+{\bf
R}_{gu}^{(2)}\cdot {\bf s}_{ug}).\nonumber
\end{eqnarray}
The coupling constants $J_i$ $(i=1-3)$ are defined as:
\begin{eqnarray*}
&&J_{1}=\frac{\alpha_T^2
V^2}{\varepsilon_F-\varepsilon} ,\ \ \ \ \ \ \ \
J_{2}=-\frac{\alpha_u\alpha_T
V^2}{\varepsilon_F-\varepsilon},\\
&&J_{3}=-\frac{\alpha_g \alpha_T
V^2}{\varepsilon_F-\varepsilon},\nonumber
\end{eqnarray*}
(see Eqs. (\ref{coeff}) for definition of the parameters $\alpha$
and $\beta$).   The levels $\bar E_\Lambda$ are renormalized in
accordance with Eq.~(\ref{halda}). The group generators forming the
$o_5$ algebra are three vectors ${\bf S}_{u}, {\bf R}_{u}, {\bf
\tilde{R}}={\bf {R}}^{(1)}_{ug}+{\bf {R}}^{(2)}_{gu}$ and the scalar
$A$:
\begin{eqnarray}\label{R-tilda}
S^{+}_{u} &=& \sqrt{2}( X^{1_{u}0_{u}}+X^{0_{u}\bar{1}_{u}}),\
S^z_{u}=X^{1_{u}1_{u}}-X^{\bar{1}_{u}\bar{1}_{u}},\nonumber\\
R^{+}_{u} &=& \sqrt{2}( X^{1_{u}S_{u}}-X^{S_{u}\bar{1}_{u}}),\
R^z_{u} =-(X^{0_{u}S_{u}}+X^{S_{u}0_{u}}), \nonumber\\
{\tilde R}^{+}&=&\sqrt{2}(X^{1_{u}S_{g}}-X^{S_{g}\bar{1}_{u}}),\
{\tilde R}^{z}=-(X^{0_{u}S_{g}}+ X^{S_{g}0_{u}}),\nonumber\\
A&=&i(X^{S_uS_g}-X^{S_gS_u}).
\end{eqnarray}
${\bf s}_{\eta\eta'}$ are the components of local spin operators
for even and odd partial waves of the band electrons,
\begin{equation}
{\bf s}_{\eta\eta'}=\frac{1}{2}\sum_{kk'}\sum_{\sigma \sigma'}
c^\dag_{\eta k\sigma} \hat{\tau}_{\sigma \sigma'}
c_{\eta'k'\sigma'},
\end{equation}
$\hat{\tau}_{\sigma \sigma'}$ are the components of the three
Pauli matrices.

The total $SO(5)$ multiplet of  width $\Delta_{\rm SO(5)}$
determines the corresponding single-channel Kondo temperature
\cite{KKA04}
\begin{equation}\label{Kondo1}
T_{K1}= \bar D \exp\left(-\frac{2}{j_1+j_2 +\sqrt{(j_1+j_2)^2
+2j_3^2}}\right),
\end{equation}
provided $\bar D \gg T_{K1}\gg\Delta_{\rm SO(5)}.$ Here
$j_i=\rho_0J_i$, $\rho_0$ is the electron density of states on the
Fermi level $\varepsilon^{}_F$. This temperature is, however not
universal: at the energy scale $D\lesssim \Delta_{\rm SO(5)}$ the
fine structure of the multiplet determines the Kondo scattering.

A two-channel orbital Kondo effect (2COKE) is possible in the
situation shown in Fig. \ref{fig_2}, where at $D < \bar D$ the two
singlet levels $\bar E_{u,g}$, renormalized in accordance with Eq.
(\ref{halda}), form an "orbital" doublet separated by a gap
$\Delta_{TS}$ from the triplet $\bar E_{Tu}$. The conditions for
such level crossing are given by Eq. (\ref{singl-gr}) in Appendix C.
In this configuration, the SW-like Hamiltonian can be written in
terms of pseudospin Pauli matrices $\vec{\cal{T}}$ and $\vec{\tau}$
as,
\begin{eqnarray}
&&H_{\mathrm{orb}}=-K_h{\cal{T}}_z+ K_{\|}{\cal{T}}_z\sum_{\sigma}\tau_{z,\sigma}\nonumber\\
&&+\frac{K_{\bot}}{2}\left({\cal{T}}^+
\sum_{\sigma}\tau^{-}_{\sigma}+{\cal{T}}^{-}\sum_{\sigma}\tau^{+}_{\sigma}\right)\label{anizotr-H}\\
&&-K_1{\cal{T}}_z\sum_{kk'\sigma}
(c^{\dagger}_{e,k\sigma}c_{e,k'\sigma}+c^{\dagger}_{o,k\sigma}c_{o,k'\sigma})
-K_2\hat{I}\sum_{\sigma}\tau_{z,\sigma}.\nonumber
\end{eqnarray}
Here
\begin{eqnarray}
\vspace{-0.5in}
&&{\cal{T}}^+=X^{S_gS_u},\ \ \ \ {\cal{T}}^-=X^{S_uS_g},\nonumber\\
&&{\cal{T}}_z=\frac{X^{S_gS_g}-X^{S_uS_u}}{2},\nonumber\\
&&\hat{I}=X^{S_gS_g}+X^{S_uS_u},\nonumber\\
&&\tau^+_{\sigma}=\sum_{kk'}
c^{\dagger}_{e,k\sigma}c_{o,k'\sigma},\ \ \
\tau^-_{\sigma}=\sum_{kk'}
c^{\dagger}_{o,k\sigma}c_{e,k'\sigma},\nonumber\\
&&\tau_{z,\sigma}=\frac{1}{2}\left(\sum_{kk'}
c^{\dagger}_{e,k\sigma}c_{e,k'\sigma}-\sum_{kk'}
c^{\dagger}_{o,k\sigma}c_{o,k'\sigma}\right),\label{Pauli-m}
\end{eqnarray}
(see Appendix C for the definition of coupling parameters). The
first term in the Hamiltonian (\ref{anizotr-H}) is an analog of the
Zeeman term in the conventional KE. Its origin is the avoided level
crossing $\bar E_{Su}-\bar E_{sg}=K_h$ arising in the course of
Haldane renormalization due to weak interchannel hybridization in
the leads.\cite{KKA04} The spin-flip processes are absent in the
singlet states. As a result the spin degeneracy is symmetry
protected and thereby we arrive at a desirable situation, where spin
projection quantum number plays the role of channel index and the
two (orbital) channels are identical. This is the orbital 2CKE in an
effective magnetic field $K_h/2$ \cite{Aff05}. The channel isotropy
is protected by spin-rotation symmetry of the singlet state.

The second stage of the RG procedure, that starts at the energy
$\sim \bar D \gg \Delta_{TS}$ includes  inter-channel Kondo
cotunneling $\sim K_\|, K_\perp$, {\it and} indirect virtual
processes via excited triplet state $|T_u\rangle$ "inherited" from
the first stage of the RG procedure derived from the Hamiltonian
(\ref{sym-so5}). The full system of scaling equations  is reduced
to the following system encoding the orbital KE and including the
parameters relevant for the scale $T^*$ characterizing the
two-channel regime:
\begin{eqnarray}
\frac{d\kappa_{\| \sigma}}{d\ln{D}}&=&-\kappa_{\bot \sigma}^2
-2\kappa^{}_h\kappa^{}_{2\sigma}-\frac{{\bar j}_{2}^2}{2}\nonumber\\
&+&\frac{\kappa_{\| \sigma}}{4} \left[3(\kappa_{\|
\uparrow}^2+\kappa_{\|
\downarrow}^2+2\kappa_h^2\right.\nonumber\\
&+&\left.4\kappa_{2\uparrow}^2+4\kappa_{2\downarrow}^2)
+\kappa_{\bot \uparrow}^2+\kappa_{\bot \downarrow}^2\right],\nonumber\\
\frac{d\kappa_{\bot \sigma}}{d\ln{D}}&=&-\kappa_{\|
\sigma}\kappa_{\bot \sigma}-
\frac{3{\bar j}_{2}^2}{4}\nonumber\\
&+&\frac{\kappa_{\bot \sigma}}{4} \left[3(\kappa_{\|
\uparrow}^2+\kappa_{\|
\downarrow}^2+2\kappa_h^2\right.\nonumber\\
&+&\left.4\kappa_{2\uparrow}^2+4\kappa_{2\downarrow}^2)
+\kappa_{\bot \uparrow}^2+\kappa_{\bot \downarrow}^2\right],\nonumber\\
\frac{d\kappa_{2 \sigma}}{d\ln{D}}&=&-\frac{\kappa_h\kappa_{\| \sigma}}{2}\nonumber\\
&+&\frac{\kappa_{2\sigma}}{4}\left[3(\kappa_{\|
\uparrow}^2+\kappa_{\|
\downarrow}^2+2\kappa_h^2\right.\nonumber\\
&+&\left.4\kappa_{2\uparrow}^2+4\kappa_{2\downarrow}^2)
+\kappa_{\bot \uparrow}^2+\kappa_{\bot
\downarrow}^2\right],\label{sc-eq}
\end{eqnarray}
with $\kappa_i=\rho_0 K_i$ $(i=\|,\bot,h,1,2)$, and
$\sigma=\uparrow,\downarrow$ are the channel indices. Here the 3rd
order terms on the RHS are retained in accordance with the general
theory of two-channel KE \cite{NoBla80}. The energy scale $T^*$ is
enhanced due to contribution from the enhanced parameter $\bar
{j}_2$. The coordinates of the corresponding fixed point are
\begin{eqnarray}
&&\kappa_{\| \uparrow}=\kappa_{\|
\downarrow}=\frac{1+\sqrt{1+2[{\bar
j}_{2}^2-10\kappa_h^2]}}{4}\approx \nonumber\\
&&\approx \frac{1}{2}
+\frac{{\bar j}_{2}^2}{4}-\frac{5}{2}\kappa_h^2,\nonumber\\
&&\kappa_{\bot \uparrow}=\kappa_{\bot \downarrow}=\sqrt{\kappa_{\|
\uparrow}^2-\kappa_h^2-0.5{\bar
j}_{2}^2},\nonumber\\
&&\kappa_{2\uparrow}=\kappa_{2\downarrow}=\frac{\kappa_h}{2}.\label{f-p-B}
\end{eqnarray}

For the case $K^{}_h=0$ (i.e., no $S_g-S_u$ intermixing) the fixed
point (\ref{f-p-B}) transforms to:
\begin{eqnarray}
\kappa_{\| \uparrow}=\kappa_{\| \downarrow}&=&
\kappa_{\bot \uparrow}=\kappa_{\bot \downarrow}=\frac{1}{2}\nonumber\\
\kappa_{1\uparrow}=\kappa_{1\downarrow}&=&
\kappa_{2\uparrow}=\kappa_{2\downarrow}=0.\label{f-p-0}
\end{eqnarray}

In order to get the Kondo temperature one should solve Eqs.
(\ref{sc-eq}) neglecting the third order terms:
\begin{eqnarray}
\frac{d\kappa_{\| \sigma}}{d\ln{D}}&=&-\kappa_{\bot \sigma}^2
-2\kappa^{}_h\kappa_{2\sigma},\nonumber\\
\frac{d\kappa_{\bot \sigma}}{d\ln{D}}&=&-\kappa_{\|
\sigma}\kappa_{\bot
\sigma},\nonumber\\
\frac{d\kappa_{1 \sigma}}{d\ln{D}}&=&0,\nonumber\\
\frac{d\kappa_{2 \sigma}}{d\ln{D}}&=&-\frac{\kappa^{}_h\kappa_{\|
\sigma}}{2}.\label{eq-T}
\end{eqnarray}
The second and fourth Eqs. (\ref{eq-T}) give:
\begin{eqnarray}
\kappa_{2\sigma}&=&\kappa_{2\sigma}^{(0)}+
\frac{\kappa^{}_h}{2}\ln{\left(\frac{\kappa_{\bot\sigma}}{\kappa_{\bot\sigma}^{(0)}}\right)},\label{j2}
\end{eqnarray}
Using Eq.(\ref{j2}) in the first Eq.(\ref{eq-T}) we get:
\begin{eqnarray}
\frac{d\kappa_{\| \sigma}}{d\ln{D}}&=&-\kappa_{\bot \sigma}^2
-2\kappa^{}_h\kappa_{2\sigma}^{(0)}-\kappa^{2}_h
\ln{\left(\frac{\kappa_{\bot\sigma}}{\kappa_{\bot\sigma}^{(0)}}\right)},\label{jpar}
\end{eqnarray}
Eqs.(\ref{jpar}) and the second Eq.(\ref{eq-T}) give:
 \begin{eqnarray}
\kappa_{\|
\sigma}^2&=&\kappa_{\bot\sigma}^2+C^2+4\kappa^{}_h\kappa_{2\sigma}^{(0)}
\ln{\left(\frac{\kappa_{\bot\sigma}}{\kappa_{\bot\sigma}^{(0)}}\right)}\nonumber\\
&+&
 \kappa^{2}_h\ln^2{\left(\frac{\kappa_{\bot\sigma}}{\kappa_{\bot\sigma}^{(0)}}\right)},\label{jpp}
  \end{eqnarray}
with $C=\sqrt{(\kappa_{\|
\sigma}^{(0)})^2-(\kappa_{\bot\sigma}^{(0)})^2}$. Using
Eq.(\ref{jpp}) in Eq.(\ref{jpar}) and neglecting the terms
proportional to $\kappa^{2}_h$ we get the Kondo temperature,
\begin{eqnarray}
T_{K\sigma}&=&\bar{D}\exp\left\{-\frac{1}{2A}\ln{\left(\frac{\kappa_{\|\sigma}+A}
{\kappa_{\|\sigma}-A}\right)}\right\},\label{T_K}
\end{eqnarray}
where $A=\sqrt{C^2-2\kappa^{}_h\kappa_{2\sigma}^{(0)}}$.
 Taking into account that
$\displaystyle{\frac{A}{\kappa_{\|\sigma}}}\ll 1$, Eq.(\ref{T_K})
can be written as:
\begin{eqnarray}
T_{K\sigma} &\approx&\bar{D}\exp\left\{-\frac{1}{2\kappa_{\|\sigma}}\right\}.\label{T-Kap}
\end{eqnarray}

In this way we arrived at the two-channel Hamiltonian with
pseudospin operator as a source of Kondo screening and spin
indices enumerating screening channels. The Zeeman operator is
relevant for the 2CKE, and its influence on the scaling behavior
in the nearest vicinity of the quantum critical point may be
described within conformal field framework \cite{Aff05}. A
peculiar feature of the orbital 2CKE described here is the
multistage renormalization of the parameters of the bare Anderson
Hamiltonian followed by an appropriate modification of the
dynamical symmetry shared by the spectrum of the doubly occupied
TQD as displayed in Fig. \ref{fig_2}.

The present scenario of orbital 2CKE implies peculiar behavior of
various observables, such as the temperature
dependence of the tunneling conductance $G(T)$.
 It is essentially
distinct from the analogous behavior of $G(T)$ in the
``conventional" spin 2CKE. First, in the weak coupling regime $T
\gg \Delta_{\rm SO(5)}$ \cite{KKA04}, Kondo co-tunneling in the
orbital 2CKE occurs according to the {\it single} channel
scenario. Two channel over-screening occurs only at the strong
coupling regime $T \ll \Delta_{\rm SO(5)}$, where the crossover to
NFL phase takes place. Second, in the orbital 2CKE, the Kondo
temperatures in the weak and strong coupling regimes
are essentially different parameters, whereas in the spin 2CKE the
two scales are nearly the same, $T^* =\alpha T_{K2}$, where $\alpha
\lesssim 1$ (see, e.g., \cite{Potok07,Aff05,Buesser11}). Therefore
the two asymptotic regimes for $G(T)$ are
\begin{equation}
G(T)/G_0 \sim \left\{
\begin{array}{ll}
                \ln^{-2} (T/T_{K1}),& T\gg T_{K1}  \\

                \frac{1}2{}+\sqrt{T/T^*},& T\ll T^* \\
              \end{array}
\right.
\end{equation}
\begin{figure}[h]\begin{center}
  \includegraphics[width=8.5cm,angle=0]{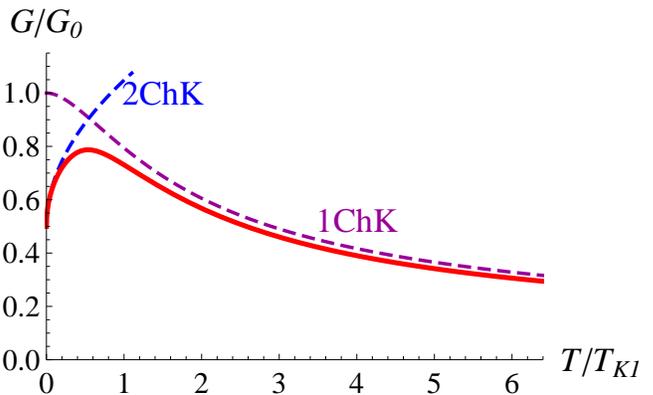}
  \caption{(Color online) Temperature dependence of tunneling conductance (solid curve).
  Low-$T$ orbital 2CKE and high-T 1CKE spin SO(5) asymptotics are shown by the dashed curves.
  The input parameters are specified in the text}
  \label{fig_3}
\end{center}
\end{figure}
This equation predicts the crossover from conventional
1CK behavior of tunneling conductance at high temperatures, to NFL
criticality characterized by $\sqrt{T}$ behavior\cite{Aff05} at low
temperatures.

 Numerical estimates of the RG  parameters using input
values of the energy intervals $\Delta_{T_g,S_u}$=0.03,
$~\Delta_{S_u,T_u}$=0.015, $\Delta_{T_u,S_g}$=0.056 (all in units of
$D_0$) show that the Haldane RG transformation stops at ${\bar
D}/D_0=0.053$ (the dependence $\bar D(\Lambda)$ is neglected in this
estimate). As is known \cite{KA01,KKA04}, the Kondo temperature for
CQD displaying $SO(n)$ symmetry is non-universal and depends on the
width of the corresponding multiplet. In our estimate
$T_{K1}(\Delta_{\rm SO(5)})$= 0.0054$D_0$. On the other hand, the
two-channel Kondo temperature $T_{K1}$ found from Eqs. (\ref{sc-eq})
without 3-rd order terms on the RHS is $T_{K2}$=0.018$D_0\gg
T_{K1}$. The reasons for such strong enhancement are (i) the
inequality $T_{K1}(\Delta_{\rm SO(5)})\ll T_{K1}(0)$ \cite{KKA04},
(ii) the contribution of the enhanced parameter $\bar {j}_2$ to the
first two equations in the system (\ref{sc-eq}): in accordance with
our numerical calculations $\bar {j}_2/j_2 \approx 1.5$, hence
$T^*/T_{K1}\gg 1$.

This result may be compared with the situation arising in doubly
occupied DQD in the serial geometry without source - drain
symmetry, where the difference between the high-energy and low-energy scales
is related to the crossover from single channel KE
to 2CKE due to asymmetry between the source/dot and drain/dot
coupling.\cite{Logan} The crossover mechanism 1CKE$\to$ 2CKE is
different in that case (two-stage Kondo screening of two spins
${\bf S}_{(s,d)}$ attached to two electrodes with essentially
differing Kondo scales ${T}_{Ks}\gg {T}_{Kd}$ \cite{Zar06}). The
1CKE is characterized by the scale ${T}_{Ks}$, and the 2CKE regime
arises at essentially lower temperature $T_{2K}\ll {T}_{Ks}$ due
to overscreening of the remaining spin 1/2  by the remaining
electrons in the source and drain leads. Comparing the two
mechanisms, we see that unlike Ref. \onlinecite{Logan}, our
mechanism leads to enhancement of the Kondo scale in the crossover
1CKE $\to$ 2CKE.
\begin{figure}[htb]
\centering
\includegraphics[width=75mm,angle=0]{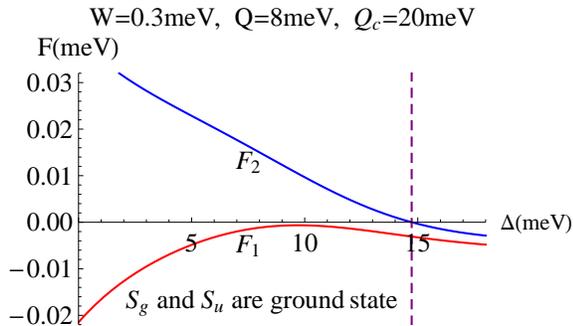}
\caption{(Color online) Functions $F_1(\Delta)$ and $F_1(\Delta)$
 defining the width of a window for orbital 2CKE [see text below and Eq. (\ref{singl-gr})
 for more details.]}
\label{b3}
\end{figure}

Experimentally, the mutual disposition of singlet and triplet levels
$E_\Lambda$ (\ref{En}) should be tunable by varying the gate
voltages which control the dot parameters such as $\Delta$, $Q_c$
and $Q$ in order to drive a TQD into the window defined by
inequalities (\ref{singl-gr}). Since these parameters enter the
corresponding equations in many ways, it is difficult to point out
their optimum combination. Figure 4 shows that the ground state of
TQD is formed by two singlets in a wide enough range of the values
of parameter $\Delta$ (marked by a vertical dashed line) at
realistic values of other model parameters.

\section{Concluding remarks}

We demonstrate in this paper that the transformation of the
conventional high-temperature 1-channel spin-Kondo effect into
unconventional low-temperature 2-channel orbital Kondo effect, where
spin enumerates the screening channels, is possible in doubly
occupied parallel TQD. The mechanism of this transformation is
related to the singlet-triplet level crossing (Fig. \ref{fig_2}).
The necessary precondition for such crossover is the presence of at
least two spin singlets in the low-energy spectrum of CQD. In case of TQD
with $N=2$ the low-energy spectrum is formed by $SO(8)$ two-triplets/two singlets
multiplet. The minimal dynamical symmetry group possessing
the demanded properties  is $SO(5)$. The theory is developed for the latter case, but the
generalization for higher $n$ is straightforward.

Specific feature of dynamical symmetry which is
crucial for 1CK $\to$ 2CK crossover is dependence of this type of
symmetry on the actual energy scale. It seen from the phase diagram
of Fig. 2 that in the ultraviolet high-energy limit $SO(5)$
dynamical symmetry is realized when the highest triplet $T_g]$ is
integrated out. The ground state in this limit is singlet, so only
the usual Kondo scattering in the weak-coupling regime due to
singlet/triplet excitations is possible. Dramatic change of Kondo
mechanism occurs because in the low-energy limit the ground state
becomes quasi doubly degenerate due to $S_u / T_u$ level crossing.
Then the roles of spin and charge degrees of freedom in Kondo
scattering swaps, and 2CK effect is realized in a strong-coupling
limit. Thus the dynamical symmetry crossover $SO(5)\to SU(2)$ is not
simple reduction of the number of relevant degrees of freedom but
reconstruction of the ground state in the course of scaling
renormalization.

Remarkable feature of this mechanism is the effect of enhancement of
the 2CK Kondo temperature in comparison with that for 1CK regime.
This effect is robust because the former "inherits" the Kondo cloud
from the latter. Enhancement of Kondo temperature accompanying
symmetry reduction $SO(5)\to SU(2)$ contrasts with the general trend
of reduction of $T_K$ with decreasing number of relevant degrees of
freedom, known, e.g. for the $SU(4)\to SU(2)$
crossover.\cite{Eto05,Lim06,Roura11}

 Several remarks about experimental
aspects of the predicted phenomenon are in order. We have already mentioned that
the most promising device for realization  of the proposed
2CKE regime within the present scenario
is an axially symmetric vertical TQD.\cite{Amaha09} This technology allows one to fabricate
symmetric or at least nearly symmetric isosceles TQD. A Small
deviation from the perfect isosceles geometry is not detrimental for
observation of the orbital 2CKE. The main effect of violating
 the left-right symmetry is expressed by the inequality $\varepsilon_l\neq
\varepsilon_r$. Inserting the corresponding corrections into Eqs.
(\ref{En}), we find that this deviation causes additional shifts of
the levels $E_{S_u}$ and $E_{S_g}$, that  eventually implies the
increment of the effective "magnetic field" $K_h$ in the Hamiltonian
$H_{\rm orb}$ (\ref{anizotr-H}). Since magnetic field is a relevant
parameter for NFL criticality,\cite{Aff05} this increase does not
prevent observation of 2CKE  as soon as the
inequalities (\ref{singl-gr}) are valid.  In
particular, the field dependence of conductance
$[G(T,K_h)-G(T,0)]/G_0$ is given by a complicated scaling function
derived in Ref. \onlinecite{PuBord04}. The quantum critical point
2CKE-1CKE may also be achieved using the mutual disposition of the
pairs $E_{T_{u,g}}$ and $E_{S_{u,g}}$ (Fig. \ref{fig_2}) governed by
the Coulomb blockade energies as control parameters. With reasonable
effort, the experimental techniques reported in Ref.
\onlinecite{Amaha09} can be modified and employed to test the
present prediction.
\ \\
\noindent {\bf Acknowledgement} This research is partially supported
by ISF grants 173/2008 and 400/12.
\appendix
\section{Eigenstates of doubly occupied symmetric TQD}\label{epsaps}

The Hamiltonian of the isolated isosceles TQD is:
\begin{eqnarray}
&&H_{d}=\sum_{a=l,r,c}\sum_{\sigma}\varepsilon_{a}d^\dagger_{a
\sigma}d_{a\sigma}\nonumber\\
&&+\sum_{a}Q_an_{a\uparrow}n_{a\downarrow}
+\sum_{i=l,r}Q_{ic}n_{i\uparrow}n_{c\downarrow}\nonumber\\
&&+W\sum_{i\sigma}(d^\dagger_{c \sigma}d_{i\sigma }+H.c.).
\label{H-dot}
\end{eqnarray}
Here $\varepsilon_a$ are the energy level positions in the central
$(c)$ and side $(l,r)$ wells of TQD, $Q_a$ and $Q_{ic}$ are
intradot and interdot Coulomb blockade parameters, respectively,
$W$ is the tunneling integral between the central and side wells.
We will be interested in the completely symmetric case,
$\varepsilon_l=\varepsilon_r\equiv \varepsilon,$ $Q_l=Q_r\equiv
Q\ll Q_c$ and $Q_{lc}=Q_{rc}\equiv Q_{ic}$.

 The Hamiltonian (\ref{H-dot}) can
be diagonalized by using the basis of two--electron wave functions
\begin{eqnarray}
&&| s_i\rangle =\frac{1}{\sqrt{2}}\left (
d^{+}_{i\uparrow}d^{+}_{c\downarrow}-d^{+}_{i\downarrow}
d^{+}_{c\uparrow}\right )\vert 0\rangle ,\nonumber \\
&&|t_i, 1\rangle = d^{+}_{i\uparrow}d^{+}_{c\uparrow}\vert
0\rangle,\nonumber\\
&&|t_i,0\rangle=\frac{1}{\sqrt{2}}
\left(d^{+}_{i\uparrow}d^{+}_{c\downarrow}+
d^{+}_{i\downarrow}d^{+}_{c\uparrow}\right )
|0\rangle ,\nonumber \\
&&|t_i, \bar{1}\rangle = d^{+}_{i\downarrow}d^{+}_{c\downarrow}
\vert 0\rangle,\nonumber\\
&&|{ex}_{i}\rangle = d^{+}_{i\uparrow}d^{+}_{i\downarrow}\vert
0\rangle,\ \ \ |{ex}_{c}\rangle =
d^{+}_{c\uparrow}d^{+}_{c\downarrow}\vert
0\rangle, \nonumber\\
&&| {ex}_{lr}\rangle =\frac{1}{\sqrt{2}}\left (
d^{+}_{l\uparrow}d^{+}_{r\downarrow}-d^{+}_{l\downarrow}
d^{+}_{r\uparrow}\right )\vert 0\rangle ,\nonumber \\
&&|{ex}_{lr}^{t}, 1\rangle = d^{+}_{l\uparrow}d^{+}_{r\uparrow}
\vert 0\rangle,\nonumber\\
&&|{ex}_{lr}^t,0\rangle=\frac{1}{\sqrt{2}}
\left(d^{+}_{l\uparrow}d^{+}_{r\downarrow}+
d^{+}_{l\downarrow}d^{+}_{r\uparrow}\right )
|0\rangle ,\nonumber \\
&&|{ex}_{lr}^t, \bar{1}\rangle =
d^{+}_{l\downarrow}d^{+}_{r\downarrow} \vert 0\rangle.
\label{basis}
\end{eqnarray}

In this basis, the Hamiltonian (\ref{H-dot}) is decomposed into
triplet and singlet matrices,
\begin{eqnarray}\label{matrices}
H_t &=& \left(
\begin{array}{ccc}
\tilde{\varepsilon} & 0 & W \\
0 & \tilde{\varepsilon} & -W \\
W & -W & 2\varepsilon \\
\end{array}\right ), \\
H_s &=& \left(
\begin{array}{cccccc}
\tilde{\varepsilon} & 0 & W & \sqrt{2}W&0&\sqrt{2}W\\
0 & \tilde{\varepsilon} & W &0&\sqrt{2}W&\sqrt{2}W\\
W & W & 2\varepsilon&0&0&0 \\
\sqrt{2}W&0&0&2\varepsilon+Q&0&0\\
0&\sqrt{2}W&0&0&2\varepsilon+Q&0\\
\sqrt{2}W&\sqrt{2}W&0&0&0&2\varepsilon_c+Q_c\\
\end{array}\right ), \nonumber
\end{eqnarray}
where $\tilde{\varepsilon}={\varepsilon}_c+\varepsilon +Q_{ic}$.

The low-energy multiplet of two-electron states found  by means of
diagonalization of the matrices (\ref{matrices}) is given by Eqs.
(\ref{En}). The eigenfunctions corresponding to the energies
(\ref{En}) are
\begin{eqnarray}
\vert S_g\rangle &=& \alpha_g\frac{\left\vert
s_l\right\rangle+\left\vert s_r\right\rangle}{\sqrt{2}}
-\beta_1|{ex}_{lr}\rangle \nonumber \\
\ \ \ \ &-&\beta_2\frac{\left\vert
{ex}_l\right\rangle+\left\vert {ex}_r\right\rangle}{\sqrt{2}} -\beta_3|{ex}_{c}\rangle,\nonumber\\
\vert T_{u}\rangle &=& \alpha^T~\frac{\left\vert t_l\right\rangle\
- \left\vert
t_r\right\rangle}{\sqrt{2}}-\beta_1|{ex}^T_{lr}\rangle,\nonumber\\
\vert S_u\rangle &=&\alpha_u \frac{\left\vert
s_l\right\rangle-\left\vert
s_r\right\rangle}{\sqrt{2}}-\beta_2\frac{\left\vert
{ex}_l\right\rangle-\left\vert {ex}_r\right\rangle}{\sqrt{2}},
\label{degen-eg-fun}\\
\vert T_{g}\rangle &=& \frac{\left\vert t_l\right\rangle +
\left\vert t_r\right\rangle}{\sqrt{2}},\nonumber
\end{eqnarray}
where
\begin{eqnarray}\label{coeff}
 \beta_1 &=& \sqrt{2}W/\Delta,
 ~\beta_2=\sqrt{2}W/(\Delta+Q),\nonumber\\
\beta_3 &=& \sqrt{2}W/(\varepsilon_c+Q_{c}-Q_{ic}-\varepsilon),
\nonumber \\
\alpha_g &=& \sqrt{1-\beta_1^2-\beta_2^2-\beta_3^2},\nonumber \\
\alpha^T &=& \sqrt{1-\beta_1^2},~ \alpha_u=\sqrt{1-\beta_2^2}.
\end{eqnarray}

Thus the lowest state of the isolated TQD is $E_{S_g}$. It is seen
from (\ref{En}) that the singlet and triplet states alternate:
$E_{S_g} < E_{T_u} < E_{S_u} < E_{T_g}$. The effective RG
procedure renormalizing the eigenvalues of quantum dot once it is
attached the leads \cite{Hald78}, has been generalized for
multilevel TQD in Ref. \onlinecite{KKA04} In accordance with this
procedure, the levels $E_\Lambda$ move downward as a function of
the scaling parameter $\eta$=$\ln (D/D_0)$ with different slopes
$\varpropto \Gamma_\Lambda$. The tunneling rates $\Gamma_\Lambda$
for even and odd states (\ref{degen-eg-fun}) obey the following
hierarchy: $\Gamma_{T_g}>\Gamma_{S_u}> \Gamma_{T_u}>\Gamma_{S_g}$.
This means that multiple level crossing is possible in the course
of RG evolution. Besides, the pairs $E_{T_{u,g}}$ and
$E_{S_{u,g}}$ are subject to level repulsion provided the left and
right tunneling channels are not completely independent.

The flow evolution of each level stops at $E_\Lambda \approx \bar
D$. The value $\bar D$ is specific for each of the four levels. At
this point the SW transformation leads to the Kondo Hamiltonian,
and the Kondo-stage of RG procedure starts in accordance with the
poor man scaling procedure.\cite{And70} The phase diagram is quite
complicated, because the two singlet and two triplet levels can
intersect in several ways depending on the values of the model
parameters $W, \tilde\varepsilon, \Delta, Q, Q_c, Q_{ic}$. Among
these scenarios are:  (I) S=1 Kondo regime corresponding to the
scenario where $T_g$ is the ground state, and the flow pattern of
dynamical symmetries with the RG procedure is $SO(8) \to SO(5) \to
SO(4) \to SO(3)$ (cf. the case of TQD with $N=4$ studied in Ref.
\onlinecite{KKA04}); (II) Absence of KE corresponding to the
scenario where  $S_g$ is the ground state; (III) Orbital KE with
almost degenerate $S_g,S_u$ ground state. Here we focus on two
competing phases, namely $SO(5)$ configuration, which involves two
singlet  states  and one triplet state, and orbital $SU(2)$
configuration, where two singlets form (quasi) degenerate pair and
the triplet state is involved as a relatively soft excitation
above this doublet.

\section{$SO(5)$-symmetry}

The effective Hamiltonians acting in the Fock space $T_u,S_g,S_u$
possess the $SO(5)$ symmetry.
 Ten group generators of the $o_5$
algebra can be combined, in particular in three  vectors and one
scalar.\cite{Dynsym} In our case these are the vectors ${\bf
S}_{u}, {\bf R}_{u}$  and the the vector intermixing $g$- and
$u$-states, namely, ${\bf \tilde{R}}={\bf {R}}^{(1)}_{ug}+{\bf
{R}}^{(2)}_{gu}$ (\ref{R-tilda}). All these operators are defined
via Hubbard operators connecting different states of the octet,
\begin{eqnarray}
S^{+}_{u} &=& \sqrt{2}( X^{1_{u}0_{u}}+X^{0_{u}\bar{1}_{u}}),\
S^{-}_{u} =(S^{+}_{u})^\dagger, \nonumber\\
S^z_{u}&=&X^{1_{u}1_{u}}-X^{\bar{1}_{u}\bar{1}_{u}},\nonumber\\
R^{+}_{u} &=& \sqrt{2}( X^{1_{u}S_{u}}-X^{S_{u}\bar{1}_{u}}),\
R^{-}_{u} =
(R^{+}_{u})^\dagger,\nonumber\\
R^z_{u} &=&-(X^{0_{u}S_{u}}+X^{S_{u}0_{u}}),  \label{comm1} \\
R^{(1)+}_{ug}&=&\sqrt{2}X^{1_{u}{S}_{g}}, \ \ \
{R}^{(1)-}_{ug}=(R^{(1)+}_{ug})^{\dag},\nonumber\\
R^{(1)z}_{ug}&=&-X^{0_{u}S_{g}},\nonumber\\
R^{(2)+}_{gu}&=&-\sqrt{2}X^{S_{g}{\bar 1}_{u}}, \ \ \ \
{R}^{(2)-}_{gu}=(R^{(2)+}_{gu})^{\dag},\nonumber\\
R^{(2)z}_{gu}&=&-X^{S_{g}0_{u}} .\nonumber
\end{eqnarray}
and the scalar operators $A$ interchanging $g,u$ variables of the
degenerate singlets:
\begin{eqnarray}\label{scal}
A&=&i(X^{S_uS_g}-X^{S_gS_u}).
\end{eqnarray}

 The exchange part of the effective $SO(5)$ Hamiltonian Hamiltonian
arising  as a result of two-stage Haldane-Anderson scaling
procedure\cite{Hald78,And70} has the form
\begin{eqnarray}
H_{spin}&=&J_{1}{\bf S}_{u}\cdot {\bf s}_u+J_{2}{\bf R}_{u}\cdot
{\bf s}_u  + J_3({\bf R}^{(1)}_{ug}\cdot {\bf s}_{gu}+{\bf
R}^{(2)}_{gu}\cdot {\bf s}_{ug})\nonumber\\
& +&J_{4}{\bf S}_{u}\cdot {\bf s}_g
+J_5{\bf \tilde{R}}\cdot {\bf s}_u \nonumber\\
&+&J_6({\bf R}_{1g}\cdot {\bf s}_{ug}+ {\bf R}_{2g}\cdot {\bf
s}_{gu})
 +J_7{\bf S}_{u}\cdot({\bf s}_{gu}+{\bf s}_{ug})\nonumber\\
 &+&
 J_8({\bf R}^{(1)}_{ug}\cdot {\bf s}_{ug}+{\bf
R}^{(2)}_{gu}\cdot {\bf s}_{gu})\nonumber\\
&+&J_{9}{\bf R}_{u}\cdot {\bf s}_g+J_{10}{\bf \tilde{R}}\cdot {\bf
s}_g+J_{11}{\bf R}_{u}\cdot({\bf s}_{gu}+{\bf s}_{ug})\nonumber\\
&+& J_{12}({\bf R}_{1g}\cdot {\bf s}_{gu}+{\bf R}_{2g}\cdot {\bf
s}_{ug}),\label{sym52-tlr}
\end{eqnarray}
where
\begin{eqnarray}
 {R}^{+}_{1g} &=&-\sqrt{2}X^{S_g\bar{1}_g},\ \ \ \
{R}^{-}_{1g}=\sqrt{2}
X^{S_g1_g}, \ \ \ \ {R}_{1gz} =-X^{S_g0_g},\nonumber\\
{R}^+_2&=&(\tilde{R}^{-}_1)^\dag,\ \ \ \ \ \ \ \ \ \;
{R}^-_2=(\tilde{R}^{+}_1)^\dag,\ \ \ \ \ \ \ \
{R}_{2z}=\tilde{R}_{1z}^\dag. \label{r-tild}
\end{eqnarray}
The spin
operators for the electrons in the leads are introduced by the
obvious relations
\begin{eqnarray}
&&{\bf s}_{g}=\frac{1}{2}\sum_{kk'}\sum_{\sigma \sigma'}
c^\dag_{gk\sigma} \hat{\tau}_{\sigma \sigma'} c_{gk'\sigma'},
\nonumber\\
&&{\bf s}_{u}=\frac{1}{2}\sum_{kk'}\sum_{\sigma \sigma'}
c^\dag_{uk\sigma} \hat{\tau}_{\sigma \sigma'} c_{uk'\sigma'},\nonumber\\
&&{\bf s}_{gu}=\frac{1}{2}\sum_{kk'}\sum_{\sigma \sigma'}
c^\dag_{gk\sigma} \hat{\tau}_{\sigma \sigma'} c_{uk'\sigma'},\ \
{\bf s}_{ug}=({\bf s}_{gu})^{\dag}.\label{gu-operators}
\end{eqnarray}
Here the first three effective exchange constants are
\begin{eqnarray}
&&J_{1}(\varepsilon)=\frac{\alpha_T^2
V^2}{\varepsilon_F-\varepsilon} ,\ \ \ \ \ \ \ \
J_{2}(\varepsilon)=-\frac{\alpha_u\alpha_T
V^2}{\varepsilon_F-\varepsilon},
\label{J}\\
&&J_{3}(\varepsilon)=-\frac{\alpha_g \alpha_T
V^2}{\varepsilon_F-\varepsilon},\nonumber
\end{eqnarray}
and the rest coupling parameters arise at the second stage of RG
procedure, which starts with the initial conditions
\begin{eqnarray}
&&J_1(\bar D)=J_{1},\ \ \ \ \ \ \ J_2(\bar
D)=J_{2},\nonumber\\
&&J_3(\bar D)=J_{3}, \ \ \ \ \ J_{i}(\bar
D)=0~~~~(i=4-12)\label{bcond5}
\end{eqnarray}

The RG flow equations for these 12 coupling constants are derived in
Ref. \onlinecite{KKA04}. Analysis of these equations shows that only
three first vertices (\ref{J}) are relevant, and one may use the
reduced Hamiltonian (\ref{sym-so5}) for calculation of the fixed
point solution, which corresponds to the Kondo temperature
\begin{eqnarray}
T_{K2}=\bar{D}\Big(1-\frac{2\sqrt{2}m_{lr}}{j_1+j_2+\sqrt{(j_1+j_2)^2+2j_3^2}}\Big)
^{\displaystyle{\frac{1}{\sqrt{2}m_{lr}}}} \label{T5-lr}
\end{eqnarray}
where $j_{i}=\rho_0 J_{i}$ ($i=1,...,12$), $d=\rho_0 D$ and
$m_{lr}=\rho_0 \bar M_{lr}.$ This expression transforms to Eq.
(\ref{Kondo1}) at $m_{lr}\to 0$.

\section{Orbital KE}\label{A3}
The singlets $S_g$ and $S_u$ become the lowest renormalized states
in the SW limit, i.e.,
$\bar{E}_{S_g}=\bar{E}_{S_u}-2{K}_{h}<\bar{E}_{T_g},\bar{E}_{T_u}$,
when
\begin{eqnarray}\label{singl-gr}
&&F_1 < 0,~~~F_2>0,\nonumber \\
\\
&&F_1 = \frac{2W^2\beta_2^2}{\beta_1^2+\beta_3^2}
\left(\frac{1}{\Delta}+\frac{2}{\varepsilon_c+Q_{c}-Q_{ic}-\varepsilon}\right)-\nonumber
\\
&&
\frac{2W^2}{\Delta+Q}+{K}_{h},\nonumber\\
&& F_2 =\frac{2W^2Q}{\Delta(\Delta+Q)}- \nonumber\\
&&\frac{2W^2(\beta_1^2-\beta_2^2)}{\beta_1^2+\beta_3^2}
\left(\frac{1}{\Delta}+\frac{2}{\varepsilon_c+Q_{c}-Q_{ic}-\varepsilon}\right)-{K}_{h}\nonumber.
\end{eqnarray}
Here $K_h/2=\bar M_{lr}$ is the indirect tunneling amplitude
between the side dots via the central dot and the leads arising in
the course of renormalization\cite{KKA04} [see also Eq.
(\ref{constants}) below].

In this case the two-channel orbital Kondo effect can be realized.
The corresponding cotunneling Hamiltonian has the form:
\begin{eqnarray}
&&H_{\mathrm{orb}}={\bar M_{lr}}(X^{uu}-X^{gg})\nonumber\\
&&+\frac{\alpha_g^2}{2}\frac{V^2}{\varepsilon_F-\varepsilon}
\sum_{kk'\sigma}X^{S_gS_g}c^{\dagger}_{e,k\sigma}c_{e,k'\sigma}\nonumber\\
&&+\frac{\alpha_u^2}{2}\frac{V^2}{\varepsilon_F-\varepsilon}
\sum_{kk'\sigma}X^{S_uS_u}c^{\dagger}_{o,k\sigma}c_{o,k'\sigma}\label{cotun-H}\\
&&+\frac{\alpha_g\alpha_u}{2}\frac{V^2}{\varepsilon_F-\varepsilon}
\sum_{kk'\sigma}(X^{S_gS_u}c^{\dagger}_{o,k\sigma}c_{e,k'\sigma}+
X^{S_uS_g}c^{\dagger}_{e,k\sigma}c_{o,k'\sigma}),\nonumber
\end{eqnarray}
with $ c_{(e,o),k\sigma}=\frac{1}{\sqrt{2}}\left(c_{l,k\sigma} \pm
c_{r,k\sigma}\right)$,
and avoided crossing of the singlet states taken into account. The
Hamiltonian (\ref{cotun-H}) can be rewritten in terms of pseudospin
Pauli matrices $\vec{\cal{T}}$ and $\vec{\tau}$. It acquires a form
given by Eq. (\ref{anizotr-H}) with following coupling constants:
\begin{eqnarray}\label{constants}
K_h&=&2{\bar M_{lr}},\nonumber\\
K_{\|}&=&\frac{\alpha_g^2+\alpha_u^2}{2}\frac{V^2}{\varepsilon_F-\varepsilon},\nonumber\\
K_{\bot}&=&\frac{\alpha_g\alpha_uV^2}{\varepsilon_F-\varepsilon},\nonumber\\
K_1&=&K_2=\frac{\alpha_g^2-\alpha_u^2}{4}\frac{V^2}{\varepsilon_F-\varepsilon}.
\end{eqnarray}
The anisotropic Hamiltonian (\ref{anizotr-H}) describes
two-channel orbital Kondo effect in an effective "magnetic field"
$K_h$.
 The coupling constant $K_h$ remains unrenormalized (because all
the terms $\sim \tau_z^2$ contribute to $K_1$), but it affects the
renormalization of the other coupling constants. The scaling
equations have the form (\ref{sc-eq}).


\begin{thebibliography}{99}
\bibitem{foot1} The singlet-triplet level crossing was discussed in Ref.
\onlinecite{MuFi97} for a heavy impurity with four electrons
captured in a three-well trap with triangular symmetry $C_{3v}$ in
the presence of the Jahn-Teller effect.
\end{thebibliography}
\end{document}